\documentclass[aps,preprintnumbers,superscriptaddress,showpacs]{revtex4}
\usepackage{epsfig}
\usepackage{psfrag}
\usepackage{amsfonts}
\usepackage{graphicx}
\usepackage{dcolumn}
\usepackage{bm}

\begin{document}

\title{Higher derivative corrections to the entropic force from holography}

\author{Zi-qiang Zhang}
\email{zhangzq@cug.edu.cn} \affiliation{School of Mathematics and
Physics, China University of Geosciences(Wuhan), Wuhan 430074,
China}

\author{Zhong-jie Luo}
\email{luozhj@cug.edu.cn} \affiliation{School of Mathematics and
Physics, China University of Geosciences(Wuhan), Wuhan 430074,
China}

\author{De-fu Hou}
\email{houdf@mail.ccnu.edu.cn} \affiliation{Key Laboratory of
Quark and Lepton Physics (MOE), Central China Normal University,
Wuhan 430079, China}

\begin{abstract}
The entropic force has been recently argued to be responsible for
dissociation of heavy quarkonia. In this paper, we analyze $R^2$
corrections and $R^4$ corrections to the entropic force,
respectively. It is shown that for $R^2$ corrections, increasing
$\lambda_{GB}$ (Gauss-Bonnet factor) leads to increasing the
entropic force. While for $R^4$ corrections, increasing $\lambda$
('t Hooft coupling) leads to decreasing the entropic force. Also,
we discuss how the entropic force changes with the shear viscosity
to entropy density ratio, $\eta/s$, at strong coupling.
\end{abstract}

\maketitle

\section{Introduction}
The experimental programs at LHC and RHIC have produced a new
state of matter so-called "strongly coupled quark-gluon plasma
(sQGP)" \cite{JA,KA,EV}. One of the main experimental signatures
for sQGP formation is dissociation of quarkonia \cite{KM}. It was
suggested earlier that the color screening is the main mechanism
responsible for this suppression \cite{TMA}. Subsequently, some
authors argued that the imaginary part of the heavy quark
potential may be a more important reason than screening
\cite{ML,AB,NB}. Recently, it was proposed by D. E. Kharzeev
\cite{DEK} that the entropic force would be responsible for
melting the quarkonia as well.

The entropic force is related to the increase of the entropy with
the separation between the constituents of the bound state. It is
an emergent force and does not describe other fundamental
interactions. Based on the second law of thermodynamics, it stems
from multiple interactions that drive the system toward the state
with a larger entropy. The entropic force was developed in
\cite{KHM} to explain the elasticity of polymer strands in rubber.
Subsequently, Verlinde argued \cite{EPV} that it would be
responsible for gravity, but this interesting idea may be
controversial (see for \cite{DC}) and will not be discussed here.
Recently, it was argued \cite{DEK} that the entropic force can
drive the dissociation process if one considers the process of
deconfinement as an entropic self-destruction. This argument is
based upon the Lattice results that show that there is a peak in
the heavy quark entropy around the crossover region of the sQGP
\cite{DKA1,DKA2,PPE}. However, it should be noticed that the
entropic force cannot be taken as a fundamental property of the
system, but it allows us to understand the behavior of complicated
microscopic systems not amenable to microscopic treatment. In this
work, we will restrict ourselves to its application in
dissociation of quarkonia in the sQGP.

AdS/CFT \cite{Maldacena:1997re,Gubser:1998bc,MadalcenaReview}, the
duality between a string theory in AdS space and a conformal field
theory in the physical space-time, has yielded many important
insights for studying different aspects of the sQGP. In this
approach, K. Hashimoto et al have carried out the entropic force
associated with a heavy quark pair for $\mathcal N=4$ SYM theory
in their seminal work \cite{KHA}. There, it is found that the peak
of the entropy near the transition point is related to the nature
of deconfinement and the growth of the entropy with the distance
can yield the entropic force. Soon after \cite{KHA},
investigations of the entropic force with respect to a moving
quarkonium appeared in \cite{KBF}. It is shown that the velocity
has the effect of increasing the entropic force thus enhancing the
quarkonia dissociation. Recently, we have studied the effect of
chemical potential on the entropic force and observed that the
chemical potential increases the entropic force implying that the
quarkonia dissociation is enhanced at finite density \cite{ZQ}.

In general, string theory contains higher derivatives corrections
due to the presence of stringy effects. Although very little is
known about the forms of higher derivative corrections in string
theory, given the vastness of the string landscape one may expect
that generic corrections do occur \cite{MRD}. As a concrete
example, type IIB string theory on $AdS_5\times S^5$ is dual to
$\mathcal N=4$ SYM theory. Using the relation
$\sqrt{\lambda}=\frac{L^2}{\alpha^\prime}$ ($L$ is the radius of
$AdS_5$ and $\alpha^\prime$ the reciprocal of the string tension),
the $\mathcal{O}(\alpha^\prime)$ expansion in type IIB string
theory becomes the $\frac{1}{\sqrt{\lambda}}$ expansion in SYM
theory. The leading order corrections in $1/\lambda$ ($R^4$
corrections) come from stringy corrections to the type IIB tree
level effective action of the form $\alpha^{\prime 3}R^4$. It was
argued \cite{ABR1,PB} that $R^4$ corrections to $\eta/s$ are
positive, consistent with the viscosity bound \cite{ABR2,RCM}. On
the other hand, curvature squared interactions (corresponding to
$R^2$ corrections) can be induced in the gravity sector in $AdS_5$
by including the world-volume action of D7 branes
\cite{ABR3,OA,OA1}. It was shown \cite{MB,MB1,YK} that in the five
dimensional gravity theories with $R^2$ corrections $\eta/s$ can
be lower than $1/(4\pi)$. Also, there are other observables or
quantities that have been studied in theories with higher
derivative corrections, see e.g. \cite{JN,KB1,JN1,ZQ2}.

In this paper, we study $R^2$ corrections and $R^4$ corrections to
the entropic force. More specially, we would like to see how these
corrections affect the enropic force as well as the quarkonia
dissociation. On the other hand, $\eta/s$ is different than
$1/(4\pi)$ in the theories with higher derivative corrections, so
the connection between $\eta/s$ and the entropic force in these
theories may be an interesting fact that comes for free in
holography. These are the main motivations of the present work.

The organization of the paper is as follows. In section 2, we
analyze $R^2$ corrections to the entropic force and explore how
these corrections affect the quarkonia dissociation. Also, we
discuss how the entropic force changes with $\eta/s$ in this case.
In section 3, we investigate $R^4$ corrections to the entropic
force as well. Finally, we provide a concluding discussion in
section 4.

\section{$R^2$ corrections}
In string theory, the $R^2$ interactions is argued to arise from
the world-volume action of D7 branes \cite{ABR3,OA,OA1}.
Restricting to the gravity sector in $AdS_5$, the effective
gravity action to leading order can be written as \cite{MB,MB1}
\begin{equation}
I=\frac{1}{16\pi G_5}\int
d^5x\sqrt{-g}[R+\frac{12}{L^2}+L^2(c_1R^2+c_2R_{\mu\nu}R^{\mu\nu}+c_3R_{\mu\nu\rho\sigma}R^{\mu\nu\rho\sigma})],\label{action}
\end{equation}
where $G_5$ is the 5 dimensional Newton constant,
$R_{\mu\nu\rho\sigma}$ is the Riemann tensor, $R$ is the Ricci
scalar, $R_{\mu\nu}$ is the Ricci tensor, $L$ is the radius of
$AdS_5$ at leading order in $c_i$ with
$\lim_{\lambda\rightarrow\infty}c_i=0$. Other terms with
additional derivatives or factors of $R$ are suppressed by higher
powers of $\frac{\alpha\prime}{L^2}$. However, at this order only
$c_3$ is unambiguous while $c_1$ and $c_2$ can be arbitrarily
altered by a field redefinition \cite{MB,MB1,YK}. To avoid this
issue, one applies the Gauss-Bonnet (GB) gravity, a special case
of the action (\ref{action}), in which $c_i$ are fixed in terms of
a single parameter $\lambda_{GB}$. The GB gravity gives the
following action \cite{BZ1}
\begin{equation}
I=\frac{1}{16\pi G_5}\int
d^5x\sqrt{-g}[R+\frac{12}{L^2}+\frac{\lambda_{GB}}{2}L^2(R^2-4R_{\mu\nu}R^{\mu\nu}+R_{\mu\nu\rho\sigma}R^{\mu\nu\rho\sigma})],\label{action1}
\end{equation}
where $\lambda_{GB}$ is constrained in
\begin{equation}
-\frac{7}{36}<\lambda_{GB}\leq\frac{9}{100},
\end{equation}
where the lower bound originates from requiring the boundary
energy density to be positive-definite \cite{DM} and the upper
bound comes from avoiding causality violation in the boundary
\cite{MB1}.

The black brane solution of GB gravity can be written as \cite{RG}
\begin{equation}
ds^2=-a^2\frac{r^2}{L^2}f(r)dt^2+{\frac{r^2}{L^2}}d\vec{x}^2+\frac{L^2}{r^2}\frac{dr^2}{f(r)}
\label{metric},
\end{equation}
with
\begin{equation}
f(r)=\frac{1}{2\lambda_{GB}}\big{[}1-\sqrt{1-4\lambda_{GB}(1-\frac{r_h^4}{r^4})}\big{]}\label{f1},
\end{equation}
and
\begin{equation}
a^2=\frac{1}{2}(1+\sqrt{1-4\lambda_{GB}}),
\end{equation}
where $\vec{x}=x_1,x_2,x_3$ represent the boundary coordinates and
$r$ denotes the coordinate of the 5th dimension. The boundary is
located at $r=\infty$. The horizon is located at $r=r_h$.
Moreover, the temperature is
\begin{equation}
T=\frac{ar_h}{\pi L^2}. \label{T}
\end{equation}

It was argued \cite{MB,MB1,YK} that
\begin{equation}
\frac{\eta}{s}=\frac{1}{4\pi}(1-4\lambda_{GB}),\label{eta}
\end{equation}
one can see that $\eta/s\geq\frac{1}{4\pi}$ can be violated for
$\lambda_{GB}>0$. And, by increasing $\lambda_{GB}$, $\eta/s$
decreases.

We now follow the calculations of \cite{KHA} to analyze the
entropic force for the background metric (\ref{metric}). The
entropic force is defined as \cite{DEK}
\begin{equation}
\mathcal{F}=T\frac{\partial S}{\partial x},\label{f}
\end{equation}
where $T$ is the temperature of the plasma, $S$ represents the
entropy, $x$ denotes the inter-quark distance.

The Nambu-Goto action is
\begin{equation}
S_{NG}=-\frac{1}{2\pi\alpha^\prime}\int d\tau d\sigma\mathcal
L=-\frac{1}{2\pi\alpha^\prime}\int d\tau
d\sigma\sqrt{-detg_{\alpha\beta}}, \label{S}
\end{equation}
with
\begin{equation}
g_{\alpha\beta}=g_{\mu\nu}\frac{\partial
X^\mu}{\partial\sigma^\alpha} \frac{\partial
X^\nu}{\partial\sigma^\beta},
\end{equation}
where $g_{\alpha\beta}$ is the induced metric and parameterized by
$(\tau,\sigma)$ on the string world-sheet. $g_{\mu\nu}$ is the
metric, $X^\mu$ is the target space coordinate.

For our purpose, we take the static gauge
\begin{equation}
t=\tau, \qquad x_1=\sigma,
\end{equation}
and assume that $r$ depends only on $\sigma$,
\begin{equation}
r=r(\sigma).
\end{equation}

Under this assumption, the lagrangian density is found to be
\begin{equation}
\mathcal L=a\sqrt{\frac{f(r)r^4}{L^4}+\dot{r}^2},\label{L}
\end{equation}
with $\dot{r}=\frac{\partial r}{\partial\sigma}$.

Note that $\mathcal L$ does not depend on $\sigma$ explicitly, so
the corresponding Hamiltonian is a constant,
\begin{equation}
\mathcal L-\frac{\partial\mathcal
L}{\partial\dot{r}}\dot{r}=constant.
\end{equation}

Imposing the boundary condition at $\sigma=0$,
\begin{equation}
\dot{r}=0,\qquad  r=r_c \qquad (r_h<r_c)\label{con},
\end{equation}
where $r=r_c$ is the deepest point of the U-shaped string.

One finds
\begin{equation}
\frac{f(r)r^4}{\sqrt{f(r)r^4+L^4\dot{r}^2}}=\sqrt{f(r_c)r_c^4},
\end{equation}
with
\begin{equation}
f(r_c)=\frac{1}{2\lambda_{GB}}\big{[}1-\sqrt{1-4\lambda_{GB}(1-\frac{r_h^4}{r_c^4})}\big{]},
\end{equation}
results in
\begin{equation}
\dot{r}=\frac{dr}{d\sigma}=\sqrt{\frac{r^4f(r)[r^4f(r)-r_c^4f(r_c)]}{L^4r_c^4f(r_c)}}\label{dotr}.
\end{equation}

By integrating (\ref{dotr}) the inter-quark distance of $Q\bar{Q}$
is obtained
\begin{equation}
x=2\int_{r_c}^{\infty}dr\sqrt{\frac{L^4r_c^4f(r_c)}{r^4f(r)[r^4f(r)-r_c^4f(r_c)]}}\label{x}.
\end{equation}

To study the effect of $R^2$ corrections on the inter-distance, we
plot $xT$ versus $\varepsilon$ with $\varepsilon\equiv r_h/r_c$
for different $\lambda_{GB}$ in the left panel of Fig.1. In the
plots from top to bottom $\lambda_{GB}=-0.1,0.01,0.06$
respectively. We can see that for each plot there exists a maximum
value of $xT$, and that $xT$ is an increasing function of
$\varepsilon$ for $xT<xT_{max}$ but a decreasing one for
$xT>xT_{max}$. In fact, for the later case, one needs to consider
some new configurations \cite{DB} which are not solutions of the
Nambu-Goto action. Here we are interested mostly in the region of
$xT<xT_{max}$. For convenience, we write $c\equiv xT_{max}$. From
the left panel of Fig.1, one also finds that increasing
$\lambda_{GB}$ leads to decreasing $c$. Therefore, one concludes
that with increasing $\lambda_{GB}$ the inter-distance decreases,
similarly to what occurred in \cite{JN1}.

\begin{figure}
\centering
\includegraphics[width=8cm]{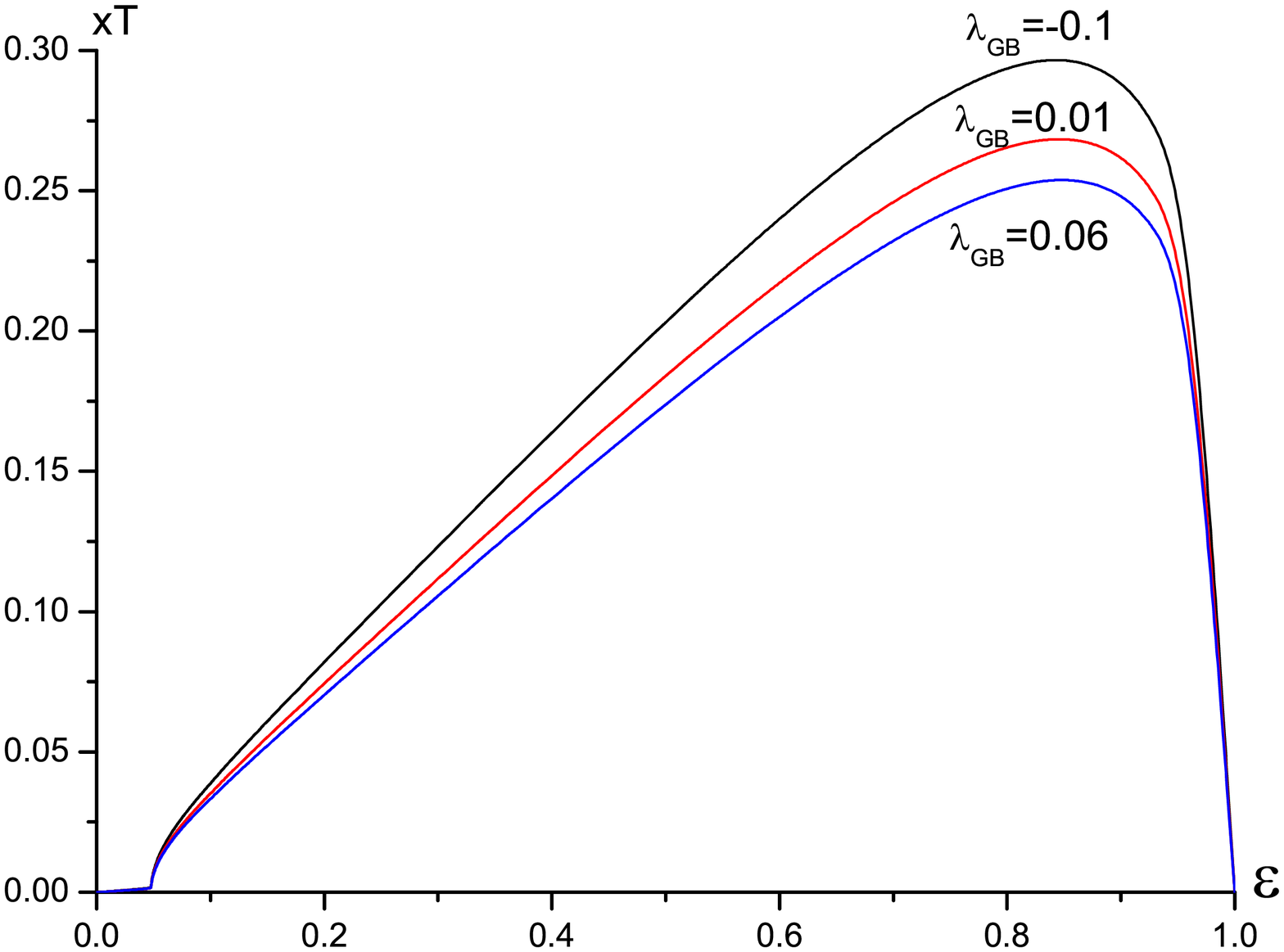}
\includegraphics[width=8cm]{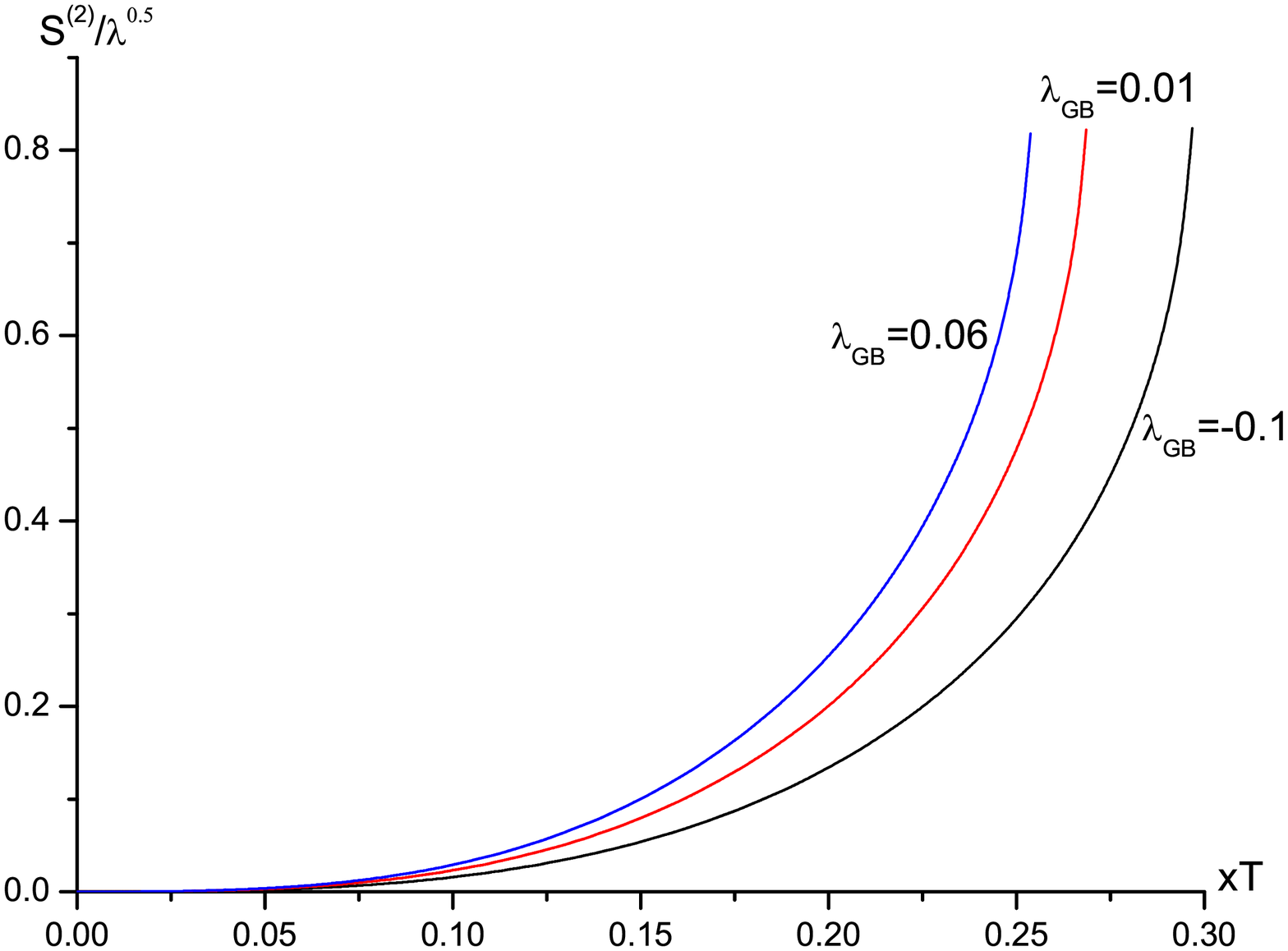}
\caption{Left: $xT$ versus $\varepsilon$ for different
$\lambda_{GB}$. From top to bottom $\lambda_{GB}=-0.1,0.01,0.06$
respectively. Right: $S^{(2)}/\sqrt{\lambda}$ versus $xT$ for
different $\lambda_{GB}$. From right to left
$\lambda_{GB}=-0.1,0.01,0.06$ respectively.}
\end{figure}

The next step is to calculate the entropy $S$, given by
\begin{equation}
S=-\frac{\partial F}{\partial T},\label{s}
\end{equation}
where $F$ is the free energy of $Q\bar{Q}$. This quantity has been
studied from the AdS/CFT correspondence, see e.g.
\cite{JMM,ABR,SJR}. There are two cases for the free energy.

1. If $x>\frac{c}{T}$, the fundamental string breaks in two
disconnected strings implying the quarks are completely screened.
In this case, the free energy is
\begin{equation}
F^{(1)}=\frac{a}{\pi\alpha^\prime}\int_{r_h}^{\infty}dr.
\end{equation}

In terms of (\ref{s}), one gets
\begin{equation}
S^{(1)}=a\sqrt{\lambda}\theta(L-\frac{c}{T})\label{S2}.
\end{equation}

2. If $x<\frac{c}{T}$, the fundamental string is connected. In
this case, the free energy is actually the total energy of the
quark pair which can be derived from the on-shell action of the
fundamental string in the dual geometry. Plugging (\ref{dotr})
into (\ref{S}), one finds
\begin{equation}
F^{(2)}=\frac{1}{\pi\alpha^\prime}\int_{r_c}^{\infty} dr
\sqrt{\frac{a^2r^4f(r)}{r^4f(r)-r_c^4f(r_c)}}.
\end{equation}

As $r_h$ is related to $T$, one can rewrite (\ref{s}) as
\begin{equation}
S=-\frac{\partial F}{\partial T}=-\frac{\partial F}{\partial
r_h}\frac{\partial r_h}{\partial T}=-\frac{\pi
L^2}{a}\frac{\partial F}{\partial r_h}.\label{s1}
\end{equation}

By virtue of (\ref{s1}), one gets
\begin{equation}
\frac{S^{(2)}}{\sqrt{\lambda}}=-\frac{1}{2a}\int_{r_c}^{\infty}
dr\frac{[a^\prime(r)b(r)+a(r)b^\prime(r)][a(r)-a(r_c)]-a(r)b(r)[a^\prime(r)-a^\prime(r_c)]}{\sqrt{a(r)b(r)[a(r)-a(r_c)]^3}}\label{S21},
\end{equation}
with
\begin{equation}
a(r)=\frac{a^2f(r)r^4}{L^4}, \qquad
a(r_c)=\frac{a^2f(r_c)r_c^4}{L^4},\qquad b(r)=a^2,
\end{equation}
where we have used the relation
$\alpha^\prime=\frac{L^2}{\sqrt{\lambda}}$. Also, the derivatives
in the above equation are with respect to $r_h$.

To proceed further we have to resort to numerical methods. In the
right panel of Fig 1, we plot $S^{(2)}/\sqrt{\lambda}$ versus $xT$
for different $\lambda_{GB}$. In the plots from right to left
$\lambda_{GB}=-0.1,0.01,0.06$, respectively. From the figures, one
can see that increasing $\lambda_{GB}$ leads to larger entropy at
small distances. On the other hand, from (\ref{f}) one finds that
the entropic force is related to the growth of the entropy with
the distance. As a result, increasing $\lambda_{GB}$ leads to
increasing the entropic force. Since the entropic force is
responsible for dissociating the quarkonia, one concludes that
increasing $\lambda_{GB}$ the quarkonia dissociation is enhanced.
Interestingly, it was argued \cite{JN1} that increasing
$\lambda_{GB}$ leads to increasing the imaginary potential thus
making the quakonia melt easier, consistent with the findings
here.

Also, it follows from (\ref{eta}) that increasing $\lambda_{GB}$
leads to decreasing $\eta/s$. While increasing $\lambda_{GB}$
leads to enhancing the quarkonia dissociation. Thus, one concludes
that in the case of $R^2$ corrections the quarkonia dissociation
is enhanced as $\eta/s$ decreases.

\section{$R^4$ corrections}
In this section we study $R^4$ corrections to the entropic force.
These corrections are related to $\alpha^\prime$ corrections on
the string theory side \cite{JP} and correspond to leading order
correction in $1/\lambda$ on the gauge theory side. The
$\alpha^\prime$-corrected metric is given by \cite{GB1}
\begin{equation}
ds^2=G_{tt}dt^2+G_{xx}d\vec{x}^2+G_{rr}dr^2 \label{metric2},
\end{equation}
with
\begin{equation}
G_{tt}=-r^2(1-w^{-4})T(w),\qquad G_{xx}=r^2X(w),\qquad
G_{rr}=r^{-2}(1-w^{-4})^{-1}U(w),
\end{equation}
where
\begin{eqnarray}
T(w)&=&1-k(75w^{-4}+\frac{1225}{16}w^{-8}-\frac{695}{16}w^{-12})+...,\nonumber\\
X(w)&=&1-\frac{25k}{16}w^{-8}(1+w^{-4})+...,\nonumber\\
U(w)&=&1+k(75w^{-4}+\frac{1175}{16}w^{-8}-\frac{4585}{16}w^{-12})+...,
\end{eqnarray}
with $w=\frac{r}{r_h}$.

The parameter $k$ is related to $\lambda$ by
\begin{equation}
k=\frac{\zeta(3)}{8}\lambda^{-3/2} \sim 0.15\lambda^{-3/2}.
\end{equation}

The horizon is $r=r_h$ and the temperature is
\begin{equation}
T=\frac{r_h}{\pi L^2(1-k)}.
\end{equation}

In addition, it was argued \cite{ABR1,PB} that
\begin{equation}
\frac{\eta}{s}=\frac{1}{4\pi}(1+\frac{135}{8}\zeta(3)\lambda^{-3/2}),\label{eta1}
\end{equation}
one can see that $\eta/s\geq1/4\pi$ remains valid in theories with
$R^4$ corrections. Also, decreasing $\lambda$ leads to increasing
$\eta/s$.

The next analysis is almost parallel to the previous section, so
we present the final results here. One finds
\begin{equation}
x=2\int_{r_c}^{\infty}dr\sqrt{\frac{a(r_c)b(r)}{a^2(r)-a(r)a(r_c)}}\label{x1},
\end{equation}
with
\begin{equation}
a(r)=r^4(1-w^{-4})T(w)X(w), \qquad
a(r_c)=r_c^4(1-w_1^{-4})T(w_1)X(w_1),\qquad b(r)=T(w)U(w),
\end{equation}
where $w_1=\frac{r_c}{r_h}$, $T(w_1)=T(w)|_{w=w_1}$,
$X(w_1)=X(w)|_{w=w_1}$.

Likewise, to analyze $R^4$ corrections to the inter-distance, we
plot $xT$ versus $\varepsilon$ with different $\lambda$ in the
left panel of Fig.2. Note that the behavior of $R^4$ corrections
is not the same as the $R^2$ corrections. Here one can see that
the value of $xT_{max}$ increases as $\lambda$ increases, in
agreement with the findings of \cite{KB1}.

On the other hand, the free energy $F^{(2)}$ is found to be
\begin{equation}
F^{(2)}=\frac{1}{\pi\alpha^\prime}\int_{r_c}^{\infty} dr
\sqrt{\frac{a(r)b(r)}{a(r)-a(r_c)}}.
\end{equation}
After some manipulations, one finds
\begin{equation}
\frac{S^{(2)}}{\sqrt{\lambda}}=\frac{k-1}{2}\int_{r_c}^{\infty}
dr\frac{[a^\prime(r)b(r)+a(r)b^\prime(r)][a(r)-a(r_c)]-a(r)b(r)[a^\prime(r)-a^\prime(r_c)]}{\sqrt{a(r)b(r)[a(r)-a(r_c)]^3}}\label{S22},
\end{equation}
with
\begin{eqnarray}
a^\prime(r)&=&r^4(T^\prime(w)X(w)+T(w)X^\prime(w)+4w^{-5}w^\prime T(w)X(w)-w^{-4}T^\prime(w)X(w)-w^{-4}T(w)X^\prime(w)),\nonumber\\
a^\prime(r_c)&=&r_c^4(T^\prime(w_1)X(w_1)+T(w_1)X^\prime(w_1)+4w_1^{-5}w_1^\prime T(w_1)X(w_1)-w_1^{-4}T^\prime(w_1)X(w_1)-w_1^{-4}T(w_1)X^\prime(w_1)),\nonumber\\
b^\prime(r)&=&T^\prime(w)U(w)+T(w)U^\prime(w),
\end{eqnarray}
and
\begin{eqnarray}
T^\prime(w)&=&300kw^{-5}w^\prime+\frac{1225}{2}kw^{-9}w^\prime-\frac{2085}{4}kw^{-13}w^\prime,\nonumber\\
T^\prime(w_1)&=&300kw_1^{-5}w_1^\prime+\frac{1225}{2}kw_1^{-9}w_1^\prime-\frac{2085}{4}kw_1^{-13}w_1^\prime,\nonumber\\
X^\prime(w)&=&\frac{25}{2}w^{-9}w^\prime+\frac{75}{4}w^{-13}w^\prime,\nonumber\\
X^\prime(w_1)&=&\frac{25}{2}w_1^{-9}w_1^\prime+\frac{75}{4}w_1^{-13}w_1^\prime,\nonumber\\
U^\prime(w)&=&-300kw^{-5}w^\prime-\frac{1175}{2}kw^{-9}w^\prime+\frac{13755}{4}kw^{-13}w^\prime,
\end{eqnarray}
where the derivatives are with respect to $r_h$.

Note that (\ref{S22}) is complicated and one needs to resort to
numerical methods. In the right panel of Fig.2, we plot
$S^{(2)}/\sqrt{\lambda}$ versus $xT$ for different $\lambda$. From
the figures, one can see that increasing $\lambda$ leads to
smaller entropy at small distances, which means the entropic force
decreases as $\lambda$ increases. In other words, decreasing
$\lambda$ enhances quarkonia dissociation. Interestingly, it was
argued that \cite{KB1} decreasing $\lambda$ leads to larger
imaginary potential or smaller dissociation length, consistent
with the findings here. Moreover, it follows from (\ref{eta1})
that decreasing $\lambda$ leads to increasing $\eta/s$. Thus, one
concludes that in the case of $R^4$ corrections the quarkonia
dissociation is enhanced as $\eta/s$ increases.

However, it should be emphasized that without computing at least
the next-to-leading order corrections in the 't Hooft coupling one
can not assure that it is physically meaningful to go all the way
down from infinity coupling to $\lambda \sim 5.5$ by just
considering $R^4$ corrections.

\begin{figure}
\centering
\includegraphics[width=8cm]{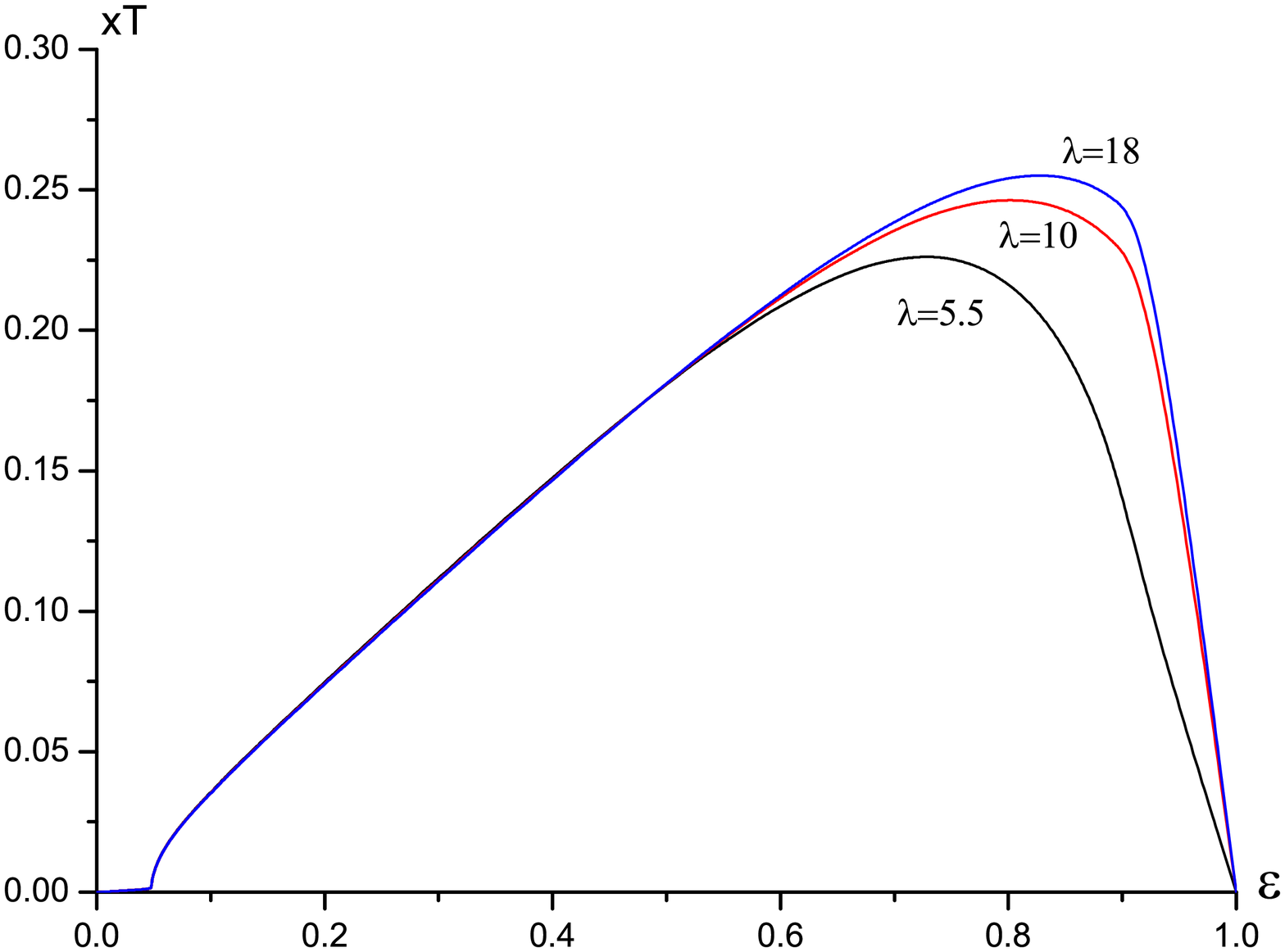}
\includegraphics[width=8cm]{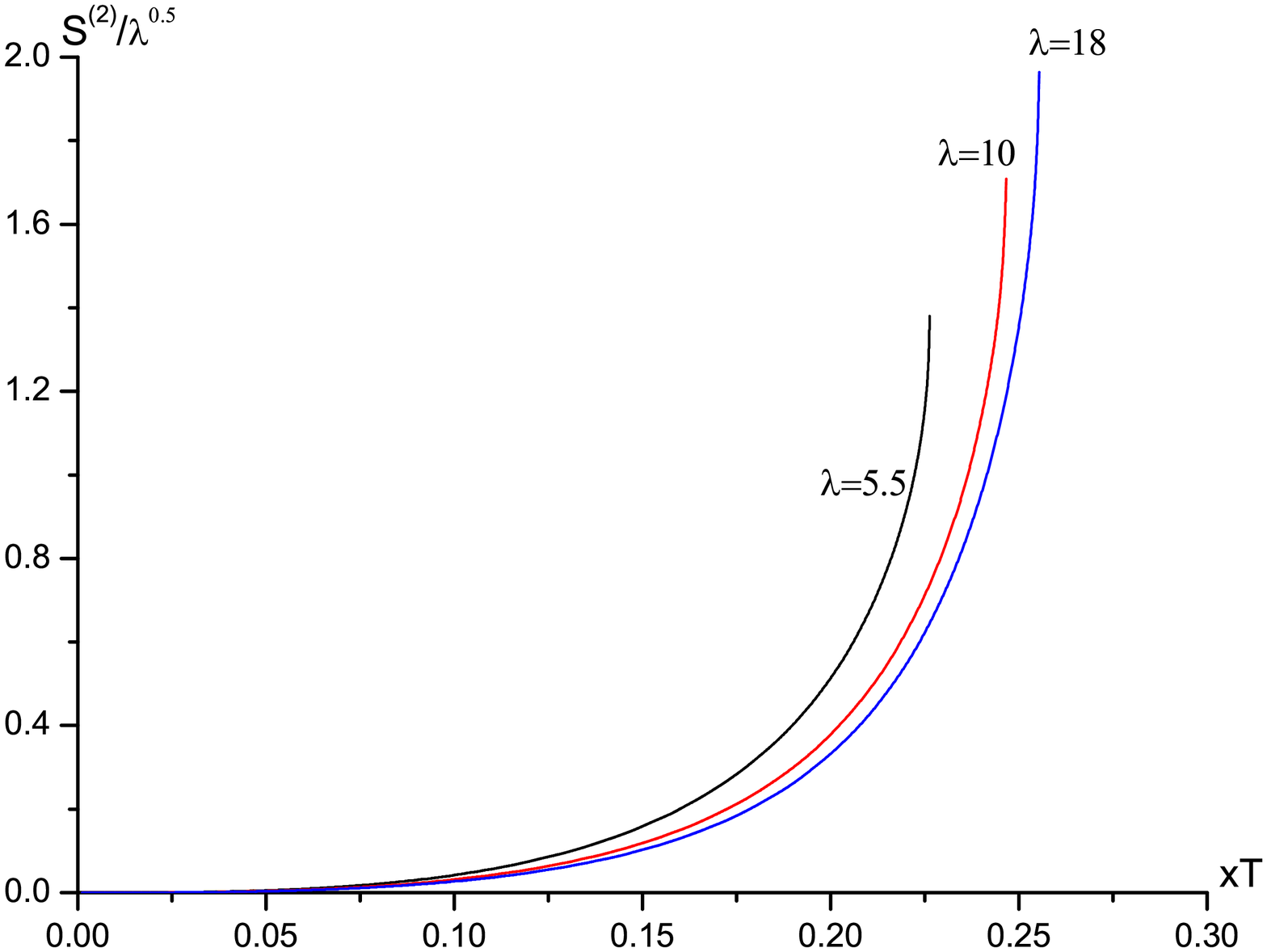}
\caption{Left: $xT$ versus $\varepsilon$ for different $\lambda$.
From top to bottom $\lambda=18,10,5.5$ respectively. Right:
$S^{(2)}/\sqrt{\lambda}$ versus $xT$ for different $\lambda$. From
right to left $\lambda=18,10,5.5$ respectively.}
\end{figure}

\section{conclusion}
The entropic force may represent a mechanism for melting the heavy
quarkonia. In this paper, we studied the effects of higher
derivative corrections to the entropic force and discussed how the
entropic force changes with $\eta/s$ at strong coupling. It is
shown that for $R^2$ corrections, increasing $\lambda_{GB}$ leads
to increasing the entropic force thus enhancing the quarkonia
dissociation. While for $R^4$ corrections, increasing $\lambda$
leads to decreasing the entropic force thus suppressing the
quarkonia dissociation. It is found that for $R^2$ corrections the
entropic force is enhanced as $\eta/s$ decreases, while for $R^4$
corrections the entropic force is enhanced as $\eta/s$ increases.
Namely, $R^2$ corrections affect the entropic force in the
opposite way of $R^4$ corrections. This is conceivable, because
$R^2$ corrections are of different nature than $R^4$ corrections
(for the origin of the two corrections, see \cite{ABR2,ABR3}). In
fact, a similar problem has been explained in the study of
$\eta/s$ \cite{ABR3}. Therein, it was argued that in certain
regimes of the parameter space, i.e., $\lambda=6\pi, N_c=3$, it is
not unreasonable to include both $R^2$ corrections and $R^4$
corrections as making independent and comparable contributions to
the CFT properties.

Certainly, one may doubt why the entropic force is the correct
(and useful) approach to understand dissociation of quarkonia.
Although we cannot provide a clear interpretation at present, we
believe that the entropic self destruction is an intriguing idea
and worth studying. Actually, in some sense one can check the
effectiveness of this idea by comparing the same effect on the
entropic force and with that on the imaginary potential. To our
knowledge, the velocity effect \cite{KBF,JN2,MAL}, the chemical
potential effect \cite{ZQ}, $R^2$ corrections \cite{JN1} and $R^4$
corrections \cite{KB1} on the two quantities give consistent
results regarding the quarkonia dissociation. These agreements
support that if the imaginary potential is right the entropic
force may be also effective.

Finally, it should be noticed that the background considered here
is a purely gravitational background with no matter fields in the
bulk and no dynamical breaking of the conformal symmetry. It is of
great interest to pursue in the investigations performed on top of
phenomenologically realistic gauge/gravity backgrounds. We hope to
report our progress in this regard in the future.

\section{Acknowledgments}
This work is partly supported by the Ministry of Science and
Technology of China (MSTC) under the ¡°973¡± Project No.
2015CB856904(4). Z-q Zhang is supported by the NSFC under Grant
No. 11705166. D-f. Hou is supported by the NSFC under Grants Nos.
11735007, 11521064.

\end{document}